# Extensions of Egghe's *g*-index: Improvements of Hirsch *h*-index

Romeo Meštrović[1], Branislav Dragović[1*]
[1]Maritime Faculty of the Univesity of Montenegro, Kotor, Montenegro
Dobrota, 85330 Kotor, Montenegro; Tel/Fax: +382 303 184
E-mails: romeo@ucg.ac.me; branod1809@gmail.com
[*]Corresponding Author, branod1809@gmail.com

## Abstract

A few new indices to characterize the scientific output of scientists are defined in the paper. These indices are compared with $h$-index and its alternative indices using some proven assertions. The $g_d$-indices are introduced as extensions of the $g$-index to define $H$-index as an improvement of the $h$-index. Numerous computational results which are conducted indicate the good behaviour of defined indices to evaluate the scientific impact of scientists. There exist very good approximations between some pairs of all considered indices in the sense that related ratios are often very close to 1.

## 1. Introduction and notations

To appraise the scientific impact of scientists, institutions and research areas among others, several publication-based indicators are used, such as the size-dependent indicators (total number of citations and number of highly cited papers) and size-independent indicators (average number of citations per paper and proportion of highly cited publications) (Waltman, 2016), as well as citation frequency (life cycle) of papers. Based on the limitations of these indicators, Hirsch (2005) proposed a new indicator called *h*-index, whilst Egghe (2006a, b, c) defined and studied an improvement of the *h*-index called the *g*-index.

The $g$-index as an improvement of the $h$-index, is defined as the highest rank such that the cumulative sum of the number of citations received is larger than or equal to the square of this rank. Notice that the $g$-index comprises information about not only the size of the productive core but also the impact of the papers in the core. Jin et al. (2007) pointed out that the $g$-index overcomes the problem that the $h$-index does not include an indicator for the internal changes of the Hirsch core. Yet, it requires drawing a longer list than necessarily for the $h$-index, hence increasing the precision problem.

Advantages and disadvantages of the *h*-index are extensive elaborated by Jin et al. (2007) and van Eck and Waltman (2008) among others, and most recently in Britoa and Rodriguez Navarro (2021) and Bihari et al. (2021), whilst the correlations between the *h*-index and more than a dozen *h*-index variants was presented by Bihari et al. (2021). For instance, the lack of sensitivity of $h$ highly-cited papers in the $h$-core (the $h$ most cited papers that are counted for $h$ because they received $h$ or more than $h$ citations) is a frequently noticed disadvantage. Notice that Rousseau (2006) states: "Certainly the $h$-index does not tell the full story, and, although a more sensitive indicator than the $h$-index, neither does the $g$-index. Taken together, $g$ and $h$ present a concise picture of a scientist's achievements in terms of publications and citations."

The $h$-index and its variants are extensively studied in the Journal of Informetrics (experimentally and theoretically). Notice that several generalizations, extensions and variations of the $h$-index and the $g$-index were defined, investigated and compared in the last two decades. For a recent review on *h*-index and its alternative indices, see Bihari et al. (2021). Schreiber (2010) considered twenty Hirsch index variants and other indicators giving more or less preference to highly cited paper. Some Hirsch-type indices were exposed by Egghe (2010). A review of the literature on citation impact indicators was presented by Waltman (2016) and by Egghe (2010).

In this paper we define and study seven new indices which are greater or equal than the $h$-index. Firstly, for a suitable positive integer $d \geq 2$, the $g_d$-indices are defined by using the condition that is stronger than that which defines $g$-index. This definition allows us to define the $\overline{g}$-, $H$-, $B$- and $F$-indices. Moreover, the $C$-, $D$- and $K$-indices are defined. We prove several inequalities involving these indices and earlier investigated $h$-, $g$-, $A$-, $R$- and $hg$-indices. These inequalities and our computational results containing all these indices for 14 Price awardees based on their citation records compiled from Scopus on February 2023, show that there exist better approximations between some pairs of these indices in the sense that related ratios are mainly very close to 1. Our computations are motivated by the fact that Glänzel and Persson (2005) calculated the $h$-index for the 14 Price awardees who are still active in quantitative studies of science (based on their published journal papers from January 1986 to August 2005 extracted by Web of Science (WoS) database).

To do it, the present paper specifies the notations and related notions as follows:

$n$ - the total number of publications of a scientist in considered database;

$cit_j$ ( $j = 1, 2, ..., n$ ) – the number of citations received by the $j$'th publication ranked in decreased order;

$N_{cit} = \sum_{j=1}^{N_w} cit_j$ - the total number of citations of a scientist in considered database;

$N_{cit}(s) = \sum_{j=1}^{s} cit_j$ ( $s = 1, 2, ..., N_p$ ) - the total number of citations of a scientist in considered database up to the rank $s = 1, 2, ..., n$;

$N_{cit}(h) = \sum_{j=1}^{h} cit_j$ - the total number of citations of a scientist in considered database into $h$-core

and

$N_{cit}(g) = \sum_{j=1}^{g} cit_j$ - the total number citations of a scientist in considered database into $g$-core.

Here, as always in the sequel, we will suppose that a scientist $S$ has published $n$ publications whose number of citations $cit_1, cit_2, cit_3 ..., cit_n$ in some database are ranked in decreased order, i.e.,

$$cit_1 \geq cit_2 \geq \cdots \geq cit_n \geq 0. \tag{1}$$

The Hirsch or $h$-index (Hirsch, 2005) is defined as the highest rank $r$ such that the first $r$ publications of a scientist received $r$ or greater than $r$ citations in considered database, or equivalently,

$$h = \max\{r : cit_r \geq r\}.$$

The Egghe's $g$-index (Egghe, 2006a,b) is defined as the maximum value of positive integers $k$ ($k = 1, 2, ..., N$) such that

$$\frac{1}{k} \sum_{j=1}^{k} cit_j \geq k. \tag{2}$$

Obviously, $h \leq g$.

Then the $g_d$-indices with $d = 2, ..., cit_1$, where $cit_1$ is a maximal number of citations of a scientist received by the $j$'th publication in considered database are introduced in Section 2. The order $r = r(g, h)$ of the $g$-index with respect to the $h$-index is defined. It follows that $g \geq g_2 \geq g_3 \geq \cdots \geq g_{r+1} \geq h$. The $H$-index is defined as $H = h\sqrt{r}$ and the $\overline{g}$-index is defined as the average of the $g, g_3, ..., g_{r+1}$ indices. Notice that the $H$-index and the $\overline{g}$-index can be considered as improvements of the $h$-index. The preliminary computational results are also given in Section 2.

According to the well-known $A$-, $R$- and $hg$-indices (Jin, 2006, Jin et al., 2007 and Alonso et al., 2010), the four new $D$-, $C$- $F$- and $K$-indices are defined in Section 3. Some inequalities involving all these indices and the $h$- and $g$-indices are proved. Numerous computational results, discussions and remarks concerning comparison of all these indices are conducted.

Finally, concluding remarks are presented in Section 4.

## 2. The $g_d$-indices, the $\overline{g}$-index and the $H$-index

Numerous computations in several publications involving the $h$-index and the $g$-index show that the values of quotients $h/g$ are often values much greater than 1, especially, greater than 1.5, and sometimes greater than 2. In order to refine these quotients, here we define the $g_d$-indices with $d = 2, 3, ..., cit_1$ and the $H$-index which presents an improvement of the $h$-index. Here, as always in the sequel, we use the notations of Section 1 and the condition (1).

### 2.1. Definitions and preliminary results

**Definition 2.1.** Let $x \in (0, cit_1]$ be a real number. Then the real function $G : (0, cit_1] \to \{1, 2, ..., n\}$ is defined as:

$$G(x) = \max\{k : k = 1, 2, ..., n\} \text{ such that}$$

$$\frac{2}{x} \cdot \frac{\sum_{j=1}^{k} cit_j}{k} \geq k + 1.$$

**Proposition 2.2.** *The function $G(x)$ has the following properties*:

(i) $G(x)$ *is a non-increasing function on the interval* $(0, cit_1]$;

(ii) $G(cit_1) = 1 = \min\{G(x) : x \in (0, cit_2]\}$;

(iii) $\max\{G(x) : x \in (0, c_1] = n$ *is attained for each* $x \in \left(0, 2\left(\sum_{j=1}^{n} cit_j\right)/(n(n+1))\right]$

*and*

(iv) $G(x)$ *is a piecewise constant function on* $(0, cit_1]$. *Namely, for all $k = 1,2,...,n-1$ it holds*

$$G(x) = k \text{ for each } x \in \left(\left(2\left(\sum_{j=1}^{k+1} cit_j\right)/((k+1)(k+2)), \ 2\left(\sum_{j=1}^{k} cit_j\right)/(k(k+1))\right]\right).$$

*Proof.* Put

$$A_k = \frac{\sum_{j=1}^{k} cit_j}{k} \quad (k = 1,2,...,n).$$

Then from the inequalities (1) it follows that

$$cit_{k+1} \leq \frac{\sum_{j=1}^{k} cit_j}{k} \quad (k = 1,2,...,n-1),$$

which immediately yields

$$A_k \geq A_{k+1} \text{ for all } k = 1,2,...,n-1, \text{ i.e.,}$$

$$A_1 \geq A_2 \geq ... \geq A_n. \tag{3}$$

Then all the properties (i) - (iv) easily follow from Definition 2.1 and the inequalities (3).

Suppose that $cit_1 \geq 2$. In this paper, we focus our attention to the values $G(x)$, where $x \in [2, cit_1]$ is a positive integer. Accordingly, we give the next discrete version of Definition 2.1, which gives the definition of the $g_d$-indices.

**Definition 2.3.** For given positive integer $d \in [2, cit_1]$, define the $g_d$-*indices* as the maximal positive integer

$$g_d = \max\{k : k = 1,2,...,n\} \text{ such that}$$

$$\frac{2}{d} \cdot \frac{\sum_{j=1}^{k} cit_j}{k} \geq k+1. \tag{4}$$

**Remarks 2.4.** Notice that $g_d = G(d)$, where $G : (0, cit_1] \to \{1,2,...,n\}$ is the function defined by Definition 2.3. Notice also that the inequality (4) is equivalent to the condition

$$\sum_{j=1}^{k} cit_j \geq k \frac{d(k+1)}{2}.$$

Then for each fixed integer $d \geq 2$, the function $s: N \rightarrow N$ defined as $s(n) = d(n+1)/2$ is increasing, convex and $s(1) = d > 1$. Hence, the function $s$ is the gracious function by Definition 4 in Woeginger (2009). Therefore, by Definition 5 in Woeginger (2009), the $g_d$-indices from Definition 2.1 coincides with the index $G[s]$ associated to the function $s(n) = d(n+1)/2$. Recall also that, if $s(n) = n = id$ is the identity function, then the corresponding index $G[id]$ coincides with the $g$-index.

Recall that the non-integer analogues of the $g$-index and the $g_d$-indices (with the small difference) were defined by van Eck and Waltman (2008, Definitions 3.2 and 3.3; also see Egghe and Rousseau, 2019).

**Remark 2.5.** If $cit_1 = cit_2 = ... = cit_h = 1$ and $cit_{h+1} = ... = cit_n = 0$, then the left hand side of the inequality (4) is $\leq 2/d \leq 1$ which is $< k+1$ for all $k = 1,...,cit_1$. Hence, if $cit_1 = 1$, then the $g_d$-indices do not exist for none $d \geq 2$.

**Proposition 2.6.** $\{g_d\}_{d=2}^{cit_1}$ is a non-increasing sequence, i.e., we have

$$g_2 \geq g_3 \geq \cdots \geq g_{cit_1} = 1 \tag{5}$$

Moreover, we have

$$g - 1 \leq g_2 \leq g. \tag{6}$$

*Proof.* The inequalities (5) and $g_{cit_1} = 1$ immediately follow from Proposition 2.2 and Definition 2.3.

It remains be proven $g - 1 \leq g_2 \leq g$. Since the condition (4) is stronger than the condition (2) concerning the $g$-index, it follows that $g_2 \leq g$. By definition of $g$-index, we have

$$\sum_{j=1}^{g} cit_j \geq g^2. \tag{7}$$

By the condition (1), we have

$$\sum_{j=1}^{g-1} cit_j \geq (g-1)cit_g,$$

whence we obtain

$$\sum_{j=1}^{g-1} cit_j = \frac{g-1}{g} \sum_{j=1}^{g-1} cit_j + \frac{1}{g} \sum_{j=1}^{g-1} cit_j \geq \frac{g-1}{g} \sum_{j=1}^{g-1} cit_j + \frac{g-1}{g} cit_g = \frac{g-1}{g} \sum_{j=1}^{g} cit_j \tag{8}$$

Substituting (7) into (8), we get

$$\sum_{j=1}^{g-1} cit_j \geq (g-1)g,$$

or equivalently,

$$\frac{1}{g-1}\sum_{j=1}^{g-1} cit_j \geq g.$$

Therefore, $g-1 \leq g_2$. This completes the proof of Proposition 2.6.

**Definition 2.7.** If only exists the $g_2$-index that is $\geq h$ or if there is none $g_d$-index ($d = 2,3,...$) that is $\geq h$, then the *order* $r = r(g,h)$ of the $g$-index with respect to the $h$-index is $r = 1$.

If the $g_2$- and $g_3$-indices exist, then the *order* $r = r(g,h)$ of the $g$-index with respect to the $h$-index is defined as the maximal positive integer $r$ such that $g_{r+1} \geq h$.

**Corollary 2.8.** *If $r = r(g,h) \geq 1$, then*

$$g \geq g_2 \geq g_3 \geq \cdots \geq g_{r+1} \geq h. \tag{9}$$

*Proof.* The proof immediately follows from Proposition 2.6 and Definition 2.7.

**Definition 2.9.** Let $r = r(g,h)$ be the order of the $g$-index with respect to the $h$-index. If $r = 1$, then the $\overline{g}$-*index* is defined as

$$\overline{g} = g. \tag{10}$$

If $r \geq 2$, then the $\overline{g}$-*index* is defined as the average

$$\overline{g} = \frac{g + \sum_{j=3}^{r+1} g_j}{r}. \tag{11}$$

**Corollary 2.10.** *For the $\overline{g}$-index there holds*

$$h \leq \overline{g} \leq g. \tag{12}$$

*Proof.* The inequalities (12) follow immediately from (11) and the inequalities (9) of Corollary 2.8.

**Definition 2.11.** If $r = r(g,h)$, then the $H$-*index* is defined as

$$H = h\sqrt{r}. \tag{13}$$

**Example 2.12.** Assume that $cit_1 = cit_2 = \cdots = cit_h = h \geq 2$, and hence, $h \geq cit_{h+1} \geq \cdots \geq cit_n \geq 0$. Then clearly, the $h$-index is equal to $h$. If we suppose that $g \geq h+1$, then put $B = \sum_{i=1}^{g-h} cit_{h+i}$. Notice that $B \leq (g-h)h$, whence it follows that

$$\frac{\sum_{j=1}^{g} cit_i}{g} = \frac{\sum_{j=1}^{h} c_{ij} + B}{g} \leq \frac{h^2 + (g-h)h}{g} = h \leq g-1.$$

Consequently, $g = h$. Moreover, we have

$$\frac{\sum_{j=1}^{h} cit_j}{h} = h < h+1,$$

and hence $g_2 \leq h - 1 = g - 1$. Using this and the inequality (6), we obtain $g_2 = g - 1$. Therefore, by Definitions 2.7, 2.9 and 2.11, it follows that $r = 1$ and $g = \overline{g} = h = H$.

**Remark 2.13.** The $H$-index defined by (13) can be considered as the improvement of the $h$-index, as it can be seen in the computational results given in Section 3.

**Proposition 2.14.** *Let $r = r(h, g) \geq 1$ be the order of the $g$-index with respect to the $h$-index, and let $\{cit_1, ..., cit_h, cit_{h+1}, ..., cit_{h+l}\}$ ($l \geq 0$) be the all citations that belong to the to the $g_{r+1}$-core if $r \geq 2$ and to the $g$-core if $r = 1$. If $r \geq 2$, then*

$$\frac{2N_{cit}(h)}{h(h+1)} - 2 < r \leq \left[\frac{2N_{cit}(h)}{h(h+l+1)}\right] - 1. \tag{14}$$

*Furthermore, if $r = 1$, then the above inequalities holds with $(h+l)$ instead of $(h+l+1)$ on the right hand side of (14), where $[x]$ denotes the greatest integer which does not exceed $[x]$.*

*Proof.* Let $r \geq 1$. By Definitions 2.3 and 2.7, we obtain

$$g_{r+2} < h.$$

Hence, by Definitions 2.3 and 2.7, the inequality (4) is not satisfied for $k = h$ and $d = r+2$. Therefore, the converse of the inequality (4) with $k = h$ and $d = r+2$ holds, i.e.,

$$\frac{2}{r+2} \cdot \frac{\sum_{j=1}^{h} cit_j}{h} < h+1,$$

whence taking $\sum_{j=1}^{h} cit_j = N_{cit}(h)$, it follows that

$$\frac{2N_{cit}(h)}{h(h+1)} - 2 < r, \tag{15}$$

i.e.,

$$r > \frac{2N_{cit}(h)}{h(h+1)} - 2, \tag{16}$$

which is in fact the left hand side of (14).

Now suppose that $r \geq 2$. Then by Definition 2.3, the inequality (4) is satisfied for $k = h+l$ and $d = r+1$. It follows that

$$\frac{2}{r+1} \cdot \frac{\sum_{j=1}^{h+l} cit_j}{h+l} \geq h+l+1. \tag{17}$$

By the inequalities (3), we have

$$A_h = \frac{N_{cit}(h)}{h} = \frac{\sum_{j=1}^{h} cit_j}{h} \geq A_{h+l} = \frac{\sum_{j=1}^{h+l} cit_j}{h+l}, \tag{18}$$

which substituting into (17) gives

$$\frac{2}{r+1} \cdot \frac{\sum_{j=1}^{h} cit_j}{h} \geq h+l+1,$$

whence taking $\sum_{j=1}^{h} cit_j = N_{cit}(h)$, it follows that

$$r \leq \frac{2N_{cit}(h)}{h(h+l+1)} - 1, \tag{19}$$

i.e.,

$$r \leq \left[\frac{2N_{cit}(h)}{h(h+l+1)}\right] - 1$$

which is in fact the right hand side of (14).

Finally, suppose that $r=1$. Then by Definition of $g$-index, the inequality (2) is satisfied for $k=h+l$ and $d=2$. It follows that

$$\frac{\sum_{j=1}^{h+l} cit_j}{h+l} \geq h+l. \tag{20}$$

Then as in the previous case, substituting the inequality

$$\frac{N_{cit}(h)}{h} \geq \frac{\sum_{j=1}^{h+l} cit_j}{h+l}$$

into (20), we obtain

$$\frac{N_{cit}(h)}{h} \geq h+l,$$

whence it follows that

$$r = 1 \leq \frac{N_{cit}(h)}{h(h+l)}$$

and hence,

$$r = 1 \leq \left[\frac{2N_{cit}(h)}{h(h+l)}\right] - 1.$$

This completes the proof of the proposition.

The inequalities (15) and (19) imply the inequality (14), which completes the proof of Proposition 2.14.

As the immediate consequences of Proposition 2.14, we obtain the following five results.

**Corollary 2.15.** *Let* $r \geq 2$. *Then under notations of Propositions* 2.14, *we have*

$$r = \left[\frac{2N_{cit}(h)}{h(h+1)}\right] - 1 = \left[\frac{2A_h}{h+1}\right] - 1 \qquad (21)$$

*Proof.* From the estimates (14) of Proposition 2.14, we obviously have

$$\left[\frac{2N_{cit}(h)}{h(h+1)}\right] - 2 < r \leq \left[\frac{2N_{cit}(h)}{h(h+l+1)}\right] - 1 \leq \left[\frac{2N_{cit}(h)}{h(h+1)}\right] - 1,$$

whence it follows the equality (21).

**Corollary 2.16.** *Let* $r \geq 2$. *Then under notations of Propositions* 2.14, *there holds*

$$\left[\frac{2A_h}{r+2}\right] - 1 < h \leq \left[\frac{2A_h}{r+1}\right] - (l+1),$$

*where* $A_h = \left(\sum_{j=1}^{h} cit_j\right)/h.$

*If* $r = 1$, *then*

$$2(h+l) \leq 2A_h < 3(h+1).$$

**Corollary 2.17.** *Let* $r \geq 2$. *Then*

$$l < \frac{h(h+1)^2}{2N - h(h+1)}.$$

*Proof.* From the estimates (15) and (19) of Proposition 2.14, we obtain

$$\frac{2N_{cit}(h)}{h(h+1)} - 2 < r \leq \frac{2N_{cit}(h)}{h(h+l+1)} - 1,$$

which implies

$$\frac{2N_{cit}(h)}{h(h+1)} - \frac{2N_{cit}(h)}{h(h+l+1)} < 1.$$

The above inequality immediately yields the inequality from Corollary 2.17.

From Definition 2.11 and Corollary 2.15 it follows the following equality.

**Corollary 2.18.** *Let* $r \geq 2$. *Then*

$$H = h\sqrt{\left[\frac{2N_{cit}(h)}{h(h+1)}\right] - 1} = h\sqrt{\left[\frac{2A_h}{h+1}\right] - 1}. \tag{22}$$

Corollary 2.19 directly implies the following estimates.

**Corollary 2.19.** *Let* $r \geq 2$. *Then*

$$\sqrt{\frac{2hN_{cit}(h)}{(h+1)} - 2h^2} < H \leq \sqrt{\frac{2hN_{cit}(h)}{(h+1)} - h^2},$$

*or equivalently,*

$$\sqrt{2h^2\left(\frac{A_h}{h+1} - 1\right)} < H \leq \sqrt{h^2\left(\frac{2A_h}{h+1} - 1\right)}. \tag{23}.$$

## 2.2. Preliminary computational results

Using Definitions, Propositions, expressions and notations from Subsection 2.1, based on data from Scopus database on 21 February 2023 for 14 Price awardees (see Tables A1 and A2 of Appendix), we obtain the following table.

**Table 1.** Price awardees data (bases on Scopus, February 2023) - Different indices and related values

| Name | Leydesdorff | Glänzel | Moed | Van Raan | Rousseau | Schubert | Martin |
|---|---|---|---|---|---|---|---|
| $g$ | 145 | 99 | 86 | 89 | 79 | 85 | 85 |
| $g_3$ | 114 | 76 | 68 | 70 | 62 | 68 | 70 |
| $g_4$ | 96 | 63 | 57 | 59 | 52 | 58 | 60 |
| $g_5$ | 83 | - | 50 | 52 | 45 | 51 | 54 |
| $g_6$ | - | - | - | - | - | 46 | 48 |
| $g_7$ | - | - | - | - | - | 42 | 45 |
| $g_8$ | - | - | - | - | - | - | 41 |
| $g_9$ | - | - | - | - | - | - | 39 |
| $h$ | 79 | 61 | 49 | 48 | 43 | 42 | 38 |
| $r$ | 4 | 3 | 4 | 4 | 4 | 6 | 8 |
| $c_h - h$ | 0 | 1 | 0 | 0 | 0 | 0 | 2 |
| $(c_h - h)/h$ | 0 | 0.017 | 0 | 0 | 0 | 0 | 0.053 |
| $g_{r+1} - h$ | 4 | 2 | 1 | 4 | 2 | 0 | 1 |
| $(g_{r+1} - h)/h$ | 0.051 | 0.033 | 0.020 | 0.083 | 0.047 | 0.000 | 0.026 |
| $h/g$ | 0.546 | 0.616 | 0.570 | 0.539 | 0.544 | 0.494 | 0.447 |
| $h\sqrt{r}/g$ | 1.090 | 1.067 | 1.140 | 1.079 | 1.089 | 1.210 | 1.264 |
| $h\sqrt{r-1}/g$ | 0.944 | 0.871 | 0.986 | 0.934 | 0.943 | 1.1050 | 1.183 |
| $H = h\sqrt{r}$ | 158 | 105.652 | 98 | 96 | 86 | 102.879 | 107.480 |
| $\bar{g}$ | 88.750 | 79.333 | 65.250 | 67.500 | 59.500 | 58.333 | 55.250 |
| $N_{cit}$ | 8053 | 11766 | 7606 | 8308 | 8053 | 7587 | 7598 |
| $N_{cit}(h)$ | 17360 | 8049 | 6351 | 6833 | 5203 | 6359 | 7048 |
| $N_{cit}(h)/h$ | 219.747 | 131.95 | 129.612 | 142.235 | 121 | 151.140 | 185.48 |
| $N_{cit}(g)$ | 21225 | 9810 | 7397 | 8042 | 6300 | 7348 | 7597 |
| $N_{cit}(g)/g$ | 146.379 | 99.691 | 86.01 | 90.360 | 79.747 | 86.447 | 89.376 |

**Table 1** – Continued. Price awardees data (bases on Scopus, February 2023) - Different indices and related values

| Name | Narin | Garfield | Braun | Small | Egghe | Ingwersen | White |
|---|---|---|---|---|---|---|---|
| $g$ | 68 | 106 | 66 | 57 | 69 | 59 | 28 |
| $g_3$ | 68 | 86 | 51 | 57 | 54 | 46 | 28 |
| $g_4$ | 59 | 74 | 42 | 57 | 46 | 39 | 28 |
| $g_5$ | 53 | 65 | 37 | 54 | 41 | 34 | 28 |
| $g_6$ | 48 | 60 | - | 50 | 36 | 31 | 27 |
| $g_7$ | 44 | 55 | - | 46 | 33 | 28 | 25 |
| $g_8$ | 42 | 51 | - | 43 | 31 | - | 23 |
| $g_9$ | 38 | 48 | - | 40 | - | - | 22 |
| $g_{10}$ | - | 45 | - | 38 | - | - | 21 |
| $g_{11}$ | - | 43 | - | 36 | - | - | 20 |
| $g_{12}$ | - | 41 | - | 34 | - | - | - |
| $g_{13}$ | - | 39 | - | - | - | - | - |
| $g_{14}$ | - | 38 | - | - | - | - | - |
| $h$ | 38 | 37 | 37 | 34 | 30 | 27 | 19 |
| $r$ | 8 | 13 | 4 | 11 | 7 | 6 | 10 |
| $c_h - h$ | 0 | 2 | 0 | 0 | 2 | 1 | 1 |
| $(c_h - h)/h$ | 0 | 0.054 | 0 | 0 | 0.067 | 0.037 | 0.053 |
| $g_{r+1} - h$ | 0 | 1 | 0 | 0 | 1 | 1 | 1 |
| $(g_{r+1} - h)/h$ | 0 | 0.027 | 0 | 0 | 0.033 | 0.037 | 0.053 |
| $h/g$ | 0.559 | 0.350 | 0.561 | 0.596 | 0.435 | 0.545 | 0.679 |
| $h\sqrt{r}/g$ | 1.581 | 1.259 | 1.121 | 1.978 | 1.150 | 1.121 | 2.251 |
| $h\sqrt{r-1}/g$ | 1.479 | 1.209 | 0.971 | 1.886 | 1.065 | 1.023 | 2.146 |
| $H = h_r = h\sqrt{r}$ | 107.480 | 133.405 | 74 | 112.765 | 79.373 | 66.136 | 63.017 |
| $\overline{g}$ | 52.500 | 57.769 | 49 | 46.545 | 44.286 | 39.500 | 25.000 |
| $N_{cit}$ | 7209 | 11515 | 5680 | 7693 | 5640 | 3606 | 2399 |
| $N_{cit}(h)$ | 6823 | 10509 | 3566 | 7471 | 3995 | 2952 | 2332 |
| $N_{cit}(h)/h$ | 179.55 | 284.03 | 96.378 | 219.735 | 133.167 | 109.333 | 122.74 |
| $N_{cit}(g)$ | 7209 | 11359 | 4373 | 7690 | 4807 | 3508 | 2399 |
| $N_{cit}(g)/g$ | 106.015 | 107.160 | 66.258 | 134.912 | 69.667 | 59.458 | 85.68 |

The data from Table 1 will be used in the next section to obtain numerous computational results.

# 3. The comparisons of the $h$-, $g$-, $A$- and the $R$-indices with new $\bar{g}$-, $H$- and $D$-indices

Using the notations from Sections 1 and 2, Jin's $A$-*index* (the name was suggested by Rousseau (2006)), introduced by Jin (Jin 2006) is defined as the average number of citations received by the publications into $h$-core, i.e.,

$$A = \frac{1}{h}\sum_{j=1}^{h} cit_j,$$

which is under our notation equal to $N_{cit}(h)/h$.

Another attempt to improve the insensitivity of the $h$-index to the number of citations to highly cited papers is the $R$-index, introduced in Jin et al. (2007. p.857). The $R$-*index* as defined as

$$R = \sqrt{\sum_{j=1}^{h} cit_j},$$

where publications, as usual, are ranked in decreasing order of the number of received citations, whereby $cit_j$ is the number of citations to the $j$'th publication and where $h$ is the $h$-index. So, as the $h$-index, also this measure takes into account the actual $cit_j$-values in the $h$-core. It is an improvement of the Jin's $A$-index. Note that

$$R = \sqrt{hA} = \sqrt{N_{cit}(h)}.$$

It is easy to show that (see Proposition 1 and Corollary in Jin et al. (2007))

$$h \leq g \leq A. \tag{24}$$

Moreover, from $R = \sqrt{hA}$ and $A \geq h$ by Corollary in Jin et. al. (2007, p. 657) it follows that

$$h \leq R. \tag{25}$$

Here we also introduce the $D$-index defined as

$$D = \sqrt{2hA - h^2} = \sqrt{2R^2 - h^2}.$$

Notice that

$$D = \sqrt{2\sum_{j=1}^{h} cit_j - h^2} = \sqrt{\sum_{j=1}^{h} cit_j + \sum_{j=1}^{h}(cit_j - h)} \geq R. \tag{26}$$

From the inequality $(A-h)^2 \geq 0$ it follows immediately that

$$D \leq A. \tag{27}$$

By (25), (26) and (27) we have the following result.

**Proposition 3.1.** *The inequalities*

$$h \leq R \leq D \leq A. \tag{28}$$

*holds.*

Recall that in Section 2 (Definitions 2.9 and 2.11) we defined the $\bar{g}$-index and the $H$-index as:

$$\bar{g} = g \quad \text{if} \quad r = 1,$$

$$\bar{g} = \frac{g + \sum_{j=3}^{r+1} g_j}{r} \quad \text{if} \quad r \geq 2.$$

and

$$H = h\sqrt{r},$$

where $r = r(g,h)$ is the order of the $g$-index with respect to the $h$-index.

By Corollary 2.10, $h \leq \bar{g} \leq g$, which together with the inequality (24) yields the following result.

**Proposition 3.2.** *The inequalities*

$$h \leq \bar{g} \leq g \leq A \tag{29}$$

*holds.*

**Proposition 3.3.** *The inequalities*

$$H \leq D \leq A \tag{30}$$

*holds.*

*Proof.* By (27), we have $D \leq A$ Note that the inequality $H \leq D$ is equivalent to the following one:

$$h \leq \frac{2A}{r+1}. \tag{31}$$

If $r = 1$, the inequality (31) becomes $h \leq A$, which is true. If $r \geq 2$ and $l = g_{r+1} - h$, then by Corollary 2.16,

$$h \leq \frac{2A}{r+1} - (l+1) \leq \frac{2A}{r+1} - 1 < \frac{2A}{r+1}.$$

This completes the proof.

**Remarks 3.4.** Notice that the equality in the inequality (29) and thus in the inequalities (28) holds in Example 2.12 with $cit_1 = cit_2 = \cdots = cit_h = h$ and $cit_{h+1} = \cdots = cit_n = 0$.

If $r \geq 2$, then from the proof of Proposition 3.3, we get

$$h \leq \frac{2A}{r+1} - 1.$$

From data of Table 1 and the above expressions we immediately obtain the following table.

**Table 2.** Price awardees data (bases on Scopus, February 2023) - The $h$-, $\overline{g}$-, $R$-, $g$-, $H$-, $D$- and $A$-indices

| Name | Leydesdorf | Glänzel | Moed | Van Raan | Rousseau | Schubert | Martin |
|---|---|---|---|---|---|---|---|
| $h$ | 79 | 61 | 49 | 48 | 43 | 42 | 38 |
| $\overline{g}$ | 88.750 | 79.333 | 65.250 | 67.500 | 59.500 | 58.333 | 55.250 |
| $R$ | 131.757 | 89.716 | 79.693 | 82.662 | 72.132 | 79.743 | 83.952 |
| $g$ | 145 | 99 | 86 | 89 | 79 | 85 | 85 |
| $H$ | 158 | 105.652 | 98 | 96 | 86 | 102.879 | 107.480 |
| $D$ | 168.757 | 111.252 | 101.494 | 106.593 | 92.504 | 104.555 | 112.481 |
| $A$ | 219.747 | 131.950 | 129.612 | 142.235 | 121 | 151.140 | 185.480 |

**Table 2 – Continued.** Price awardees data (bases on Scopus, February 2023) - The $h$-, $\overline{g}$-, $R$-, $g$-, $H$-, $D$- and $A$-indices

| Name | Narin | Garfield | Braun | Small | Egghe | Ingwersen | White |
|---|---|---|---|---|---|---|---|
| $h$ | 38 | 37 | 37 | 34 | 30 | 27 | 19 |
| $\overline{g}$ | 52.500 | 57.769 | 49 | 46.545 | 44.286 | 39.500 | 25.000 |
| $R$ | 82.6015 | 102.513 | 59.716 | 86.435 | 63.206 | 54.332 | 48.291 |
| $g$ | 68 | 106 | 66 | 57 | 69 | 59 | 28 |
| $H$ | 107.480 | 133.405 | 74 | 112.765 | 79.373 | 66.136 | 63.017 |
| $D$ | 110.463 | 140.175 | 75.914 | 117.414 | 84.202 | 71.938 | 65.597 |
| $A$ | 179.550 | 284.030 | 96.378 | 219.735 | 133.167 | 109.330 | 122.740 |

From Table 2 we see that for 11 scientists there holds

$$h < \overline{g} < R < g < H < D < A,$$

while for other three scientists (Narin, Small and White) we have

$$h < \overline{g} < R < H < D < A$$

and for these three scientists we have.

$$R > g.$$

Notice that the order $r = r(g, h)$ of the $g$-index with respect to the $h$-index of Narin, Small and White are relatively big numbers; namely, equals 8, 11 and 10, respectively (see Table 1).

**Remarks 3.5.** Table 2 and many additional computations concerning numerous scientists assumed from different databases show that $H \geq g$. This inequality is equivalent to the following one:

$$h\sqrt{r} \geq g. \tag{32}$$

However, let $cit_1 = cit_2 = 4$, $cit_3 = 1$ and $cit_i = 0$ for $i > 3$. Then $h = 2$, $g = 3$, $g_2 = 2$ and $g_3 = 1 < h$. Thus, $r = 1$, $\overline{g} = 3$ and $H = h = 2 < g$. This shows that the inequality (32) is not necessarily satisfied for $r = 1$. Assume now that $cit_1 = cit_2 = 5$, $cit_3 = 2$ and $cit_i = 0$ for $i > 3$. Then $h = 2$, $g = 3$, $g_2 = 3$, $g_3 = 2$ and $g_4 = 1 < h$. Thus, $r = 2$, $\overline{g} = 2.5$ and $H = h\sqrt{r} = 2\sqrt{2} < 3 = g$. This shows that the inequality (32) is not necessarily satisfied for $r = 2$. Similarly, we can construct an example with arbitrary integer $r \geq 3$ such that the converse of the inequality (32) is true.

From Table 2 we obtain the following table.

**Table 3.** Price awardees data (bases on Scopus, February 2023) - The ratios $\overline{g}/h$, $R/\overline{g}$, $g/R$, $H/g$, $D/H$, $A/D$, $A/h$ and $2A/((r+1)h)$ based on data from Table 2

| Name | Leydesdorf | Glänzel | Moed | Van Raan | Rousseau | Schubert | Martin |
|---|---|---|---|---|---|---|---|
| $\overline{g}/h$ | 1.123 | 1.301 | 1.331 | 1.406 | 1.383 | 1389 | 1.454 |
| $R/\overline{g}$ | 1.485 | 1.131 | 1.221 | 1.225 | 1.212 | 1.367 | 1.519 |
| $g/R$ | 1.101 | 1.103 | 1.079 | 1.077 | 1.095 | 1.066 | 1.012 |
| $H/g$ | 1.090 | 1.067 | 1.140 | 1.079 | 1.090 | 1.210 | 1.264 |
| $D/H$ | 1.068 | 1.053 | 1.036 | 1.103 | 1.076 | 1.016 | 1.047 |
| $A/D$ | 1.302 | 1.186 | 1.271 | 1.334 | 1.308 | 1.446 | 1.649 |
| $A/h$ | 2.782 | 2.163 | 2.645 | 2.963 | 2.813 | 3.599 | 4.881 |
| $2A/((r+1)h)$ | 1.113 | 1.082 | 1.058 | 1.185 | 1.126 | 1.028 | 1.085 |

**Table 3 – Continued.** Price awardees data (bases on Scopus, February 2023) - The ratios $\overline{g}/h$, $R/\overline{g}$, $g/R$, $H/g$, $D/H$, $A/D$, $A/h$ and $2A/((r+1)h)$ based on data from Table 2

| Name | Narin | Garfield | Braun | Small | Egghe | Ingwersen | White |
|---|---|---|---|---|---|---|---|
| $\overline{g}/h$ | 1.382 | 1.561 | 1.324 | 1.552 | 1.476 | 1.463 | 1.316 |
| $R/\overline{g}$ | 1.573 | 1.775 | 1.219 | 1.857 | 1.427 | 1.375 | 0.580 |
| $g/R$ | **0.823** | 1.034 | 1.105 | **0.659** | 1.092 | 1.086 | **0.580** |
| $H/g$ | 1.581 | 1.259 | 1.212 | 1.978 | 1.150 | 1.121 | 2.251 |
| $D/H$ | 1.028 | 1.051 | 1.026 | 1.041 | 1.061 | 1.088 | 1.041 |
| $A/D$ | 1.625 | 2.026 | 1.267 | 1.871 | 1.582 | 1.520 | 1.871 |
| $A/h$ | 9.450 | 15.352 | 5.206 | 12.926 | 8.878 | 8.100 | 12.920 |
| $2A/((r+1)h)$ | 1.050 | 1.097 | 1.042 | 1.077 | 1.110 | 1.157 | 1.175 |

**Remarks 3.6.** From Table 3 we see the ratios $H/g = h\sqrt{r}/g$ of White, Small and Narin are significantly greater than 1, namely, they are equal to 2.251, 1.978 and 1.581, respectively. Notice that the orders $r = r(g,h)$ of the $g$-index with respect to the $h$-index of White, Small and Narin are relatively big numbers; namely, equals 10, 11 and 8, respectively (see Table 1). For other 11 scientists, we have $1 < H/g < 1.3.$ Furthermore, from Table 3 we also see that the ratios $A/h$ of these three scientists are relatively big numbers, namely, they are equal to 9.450, 12.926 and 12.920, respectively.

From Table 2 we can rank Price awardees with respect to their $h$-, $\bar{g}$-, $R$-, $g$-, $H$-, $D$- and $A$-indices, as it is given in Table 4.

**Table 4.** Price awardees ranked with respect to the $h$-, $\bar{g}$-, $R$-, $g$-, $H$-, $D$- and $A$-indices

| Name | Leydesdorf | Glänzel | Moed | Van Raan | Rousseau | Schubert | Martin |
|---|---|---|---|---|---|---|---|
| $h$ | 1 | 2 | 3 | 4 | 5 | 6 | 7/8 |
| $\bar{g}$ | 1 | 2 | 4 | 3 | 5 | 6 | 8 |
| $R$ | 1 | 3 | 8 | 6 | 10 | 9 | 5 |
| $g$ | 1 | 3 | 5 | 4 | 8 | 6/7 | 6/7 |
| $H$ | 1 | 6 | 8 | 9 | 10 | 7 | 4/5 |
| $D$ | 1 | 5 | 9 | 7 | 10 | 8 | 4 |
| $A$ | 2 | 9 | 10 | 7 | 13 | 6 | 4 |

**Table 4 – Continued.** Price awardees ranked with respect to the $h$-, $\bar{g}$-, $R$-, $g$-, $H$-, $D$- and $A$-indices

| Name | Narin | Garfield | Braun | Small | Egghe | Ingwersen | White |
|---|---|---|---|---|---|---|---|
| $h$ | 7/8 | 9/10 | 9/10 | 11 | 12 | 13 | 14 |
| $\bar{g}$ | 9 | 7 | 10 | 11 | 12 | 13 | 14 |
| $R$ | 7 | 2 | 12 | 4 | 11 | 13 | 14 |
| $g$ | 10 | 2 | 11 | 13 | 9 | 12 | 14 |
| $H$ | 4/5 | 2 | 12 | 3 | 11 | 13 | 14 |
| $D$ | 6 | 2 | 12 | 3 | 11 | 13 | 14 |
| $A$ | 5 | 1 | 14 | 3 | 8 | 11 | 12 |

From Table 4 we see that considered scientists are ranked with respect to the $\bar{g}$-index very close as with respect to the $h$-index. Namely, five scientists have the same rank with respect to these two indices. Martin and Narin have the same $h$-index, and they are ranked 7[th] - 8[th] place with respect to the $h$-index, and 8[th] and 9[th] place with respect to the $g$-index, respectively. Moreover, Garfield and Braun have the same $h$-index, and they are ranked 9[th] - 10[th] place with respect to the $h$-index, and 7[th] and 10[th] place with respect to the $g$-index, respectively. Moed and Van Raan are ranked 3[th] and 4[th] place with respect to the $h$-index, and 4[th] and 3[th] place with respect to the $g$-index, respectively.

Let $\alpha$ and $\beta$ be the ratios given as

$$\alpha = \frac{H}{g} = \frac{h\sqrt{r}}{g},$$

$$\beta = \frac{D}{g} = \frac{\sqrt{2hA - h^2}}{g} = \frac{\sqrt{2R^2 - h^2}}{g}.$$

Then we have

$$\frac{D}{H} = \frac{\beta}{\alpha} = \frac{\sqrt{2hA - h^2}}{h\sqrt{r}} = \frac{\sqrt{2R^2 - h^2}}{h\sqrt{r}} = \sqrt{\frac{2A - h}{hr}}.$$

We also define the difference $\delta$ as

$$\delta = \left[\frac{2A}{h}\right] - (r+1) = \left[\frac{2N_{cit}(h)}{h^2}\right] - (r+1)$$

and

$$l = g_{r+1} - h.$$

From Tables 1 and 2, we immediately obtain the following table.

**Table 5.** Price awardees - The values $\beta, r+1, \delta$ and $l$

| Name | Leydesdorff | Glänzel | Moed | Van Raan | Rousseau | Schubert | Martin |
|---|---|---|---|---|---|---|---|
| $\beta = D/g$ | 1.164 | 1.124 | 1.180 | 1.198 | 1.171 | 1.230 | 1.323 |
| $r+1$ | 5 | 4 | 5 | 5 | 5 | 7 | 9 |
| $\delta = [2A/h] - (r+1)$ | 0 | 0 | 0 | 0 | 0 | 0 | 0 |
| $l = g_{r+1} - h$ | 4 | 2 | 1 | 4 | 2 | 1 | 1 |

**Table 5 – Continued.** Price awardees - The values $\beta, r+1, \delta$ and $l$

| Name | Narin | Garfield | Braun | Small | Egghe | Ingwersen | White |
|---|---|---|---|---|---|---|---|
| $\beta = D/g$ | 1.624 | 1.322 | 1.163 | 2.060 | 1.222 | 1.219 | 2.339 |
| $r+1$ | 9 | 14 | 5 | 12 | 8 | 7 | 11 |
| $\delta = [2A/h] - (r+1)$ | 0 | 1 | 0 | 0 | 0 | 0 | 1 |
| $l = g_{r+1} - h$ | 0 | 1 | 0 | 0 | 1 | 1 | 1 |

**Remarks 3.7.** From Tables 1 and 5 we see that for all 14 scientists, we have

$$h < \frac{2A}{r+1} < 1.2h.$$

From Table 5 we see that $r = [2A/h] - 1$ for 12 scientists, while $r = [2A/h] - 2$ for two scientists (Garfield and White).

**Remarks 3.8.** Notice that by (21) of Corollary 2.15, we have

$$r = \left[\frac{2A}{h+1}\right] \tag{33}$$

Finally, motivated by the fact that the $R$-index ($R = \sqrt{hA}$) is an improvement of the $h$-index, we believe that it can be of interest to investigate the following indices:

$$B =: \sqrt{hg}, \ C =: \sqrt{hR} = \sqrt[4]{h^3 A}, \ E =: \sqrt{hg}, \ F =: \sqrt{hH} = h\sqrt[4]{r} \text{ and } K =: \sqrt{hD} = h\sqrt{\frac{2A}{h} - 1}. \quad (34)$$

Alonso et al. (2010) presented a new index, called $hg$-index (here named the $E$-index) defined as

$$E = \sqrt{hg}.$$

Then obviously, $h \leq E \leq g$, which together with the inequality (24) yields

$$h \leq E \leq g \leq A.$$

From the above inequality, Propositions 3.1, 3.2 and 3.3 we obtain the following inequalities, respectively.

**Proposition 3.9.** *The following inequalities are satisfied*:

$$h \leq C \leq K \leq A, \quad (35)$$

$$h \leq B \leq E \leq g \leq A, \quad (36)$$

and

$$F \leq K \leq A. \quad (37)$$

From Table 2 and the expressions given by (34), we get the following table.

**Table 6.** Price awardees - The $h$-, $B$-, $C$-, $E$-, $F$- and $K$-indices

| Name | Leydesdorf | Glänzel | Moed | Van Raan | Rousseau | Schubert | Martin |
|---|---|---|---|---|---|---|---|
| $h$ | 79 | 61 | 49 | 48 | 43 | 42 | 38 |
| $B$ | 83.734 | 69.565 | 56.544 | 56.921 | 50.582 | 49.497 | 45.820 |
| $C$ | 102.024 | 73.979 | 62.490 | 62.990 | 55.693 | 57.872 | 56.482 |
| $E$ | 107.028 | 77.711 | 64.915 | 65.361 | 58.284 | 59.750 | 56.833 |
| $F$ | 111.723 | 80.279 | 69.297 | 67.882 | 60.811 | 65.734 | 63.908 |
| $K$ | 115.463 | 82.379 | 70.521 | 71.530 | 63.069 | 66.267 | 65.378 |

**Table 6 - Continued.** Price awardees - The $h$-, $B$-, $C$-, $E$-, $F$- and $K$-indices

| Name | Narin | Garfield | Braun | Small | Egghe | Ingwersen | White |
|---|---|---|---|---|---|---|---|
| $h$ | 38 | 37 | 37 | 34 | 30 | 27 | 19 |
| $B$ | 44.665 | 46.233 | 42.579 | 39.781 | 36.450 | 32.657 | 21.795 |
| $C$ | **56.026** | 61.587 | 47.005 | **54.211** | 43.545 | 38.301 | **30.291** |
| $E$ | **50.833** | 62.626 | 49.417 | **44.023** | 45.497 | 39.912 | **23.065** |
| $F$ | 63.908 | 70.257 | 52.326 | 61.919 | 48.797 | 42.257 | 34.602 |
| $K$ | 64.789 | 72.017 | 52.998 | 63.183 | 50.260 | 44.072 | 35.304 |

Using the equalities given by (34) and data from Table 6, we obtain the following table.

**Table 7.** Price awardees - The ratios $B/h$, $C/B$, $E/C$, $F/E$ and $K/F$

| Name | Leydesdorf | Glänzel | Moed | Van Raan | Rousseau | Schubert | Martin |
|---|---|---|---|---|---|---|---|
| $B/h$ | 1.060 | 1.140 | 1.110 | 1.186 | 1.176 | 1.179 | 1.206 |
| $C/B$ | 1.218 | 1.063 | 1.106 | 1.107 | 1.101 | 1.169 | 1.233 |
| $E/C$ | 1.049 | 1.050 | 1.039 | 1.038 | 1.047 | 1.032 | 1.007 |
| $F/E$ | 1.044 | 1.033 | 1.068 | 1,040 | 1.043 | 1.100 | 1.124 |
| $K/F$ | 1.033 | 1.026 | 1.018 | 1.0537 | 1.037 | 1.008 | 1.023 |

**Table 7.** – **Continued.** Price awardees - The ratios $B/h$, $C/B$, $E/C$, $F/E$ and $K/F$

| Name | Narin | Garfield | Braun | Small | Egghe | Ingwersen | White |
|---|---|---|---|---|---|---|---|
| $B/h$ | 1.223 | 1.250 | 1.151 | 1.170 | 1.215 | 1.210 | 1.147 |
| $C/B$ | 1.254 | 1.332 | 1.104 | 1.363 | 1.195 | 1.173 | 1.3401. |
| $E/C$ | 0.907 | 1.017 | 1.051 | 0.812 | 1.045 | 1.042 | 0.761 |
| $F/E$ | 1.257 | 1.122 | 1.059 | 1.407 | 1.073 | 1.059 | 1.500 |
| $K/F$ | 1.014 | 1.025 | 1.013 | 1.020 | 1.030 | 1.043 | 1.020 |

From Table 6 we see that for 11 scientists the following chain of inequalities is satisfied:

$$h < B < C < E < F < K,$$

while for other three scientists (Narin, Small and White) we have

$$h < B < F < K,$$

and for these three scientists we have.

$$C > E.$$

Furthermore, from Table 7 we also see that the values $K/F$ of 14 scientists are very close to 0 (namely, less than 1.06). The same estimate is true for the quotients $E/C$ for eleven scientists (except Narin, Small and White).

By using data from Table 6, we can rank all 14 scientists with respect to their $B$-, $C$-, $E$-, $F$- and $K$-indices, as it was given in Table 8.

**Table 8.** Price awardees ranked with respect to the $h$-, $B$-, $C$-, $E$-, $F$- and $K$-indices

| Name | Leydesdorf | Glänzel | Moed | Van Raan | Rousseau | Schubert | Martin |
|---|---|---|---|---|---|---|---|
| $h$ | 1 | 2 | 3 | 4 | 5 | 6 | 7 |
| $B$ | 1 | 2 | 4 | 3 | 5 | 6 | 8 |
| $C$ | 1 | 2 | 4 | 3 | 9 | 6 | 7 |
| $E$ | 1 | 2 | 4 | 3 | 7 | 6 | 8 |
| $F$ | 1 | 2 | 4 | 5 | 10 | 6 | 7/8 |
| $K$ | 1 | 2 | 5 | 4 | 10 | 6 | 7 |

**Table 8** – Continued. Price awardees ranked with respect to the $h$-, $B$-, $C$-, $E$-, $F$- and $K$-indices

| Name | Narin | Garfield | Braun | Small | Egghe | Ingwersen | White |
|---|---|---|---|---|---|---|---|
| $h$ | 8 | 9 | 10 | 11 | 12 | 13 | 14 |
| $B$ | 9 | 7 | 10 | 11 | 12 | 13 | 14 |
| $C$ | 8 | 5 | 11 | 10 | 12 | 13 | 14 |
| $E$ | 9 | 5 | 10 | 12 | 11 | 13 | 14 |
| $F$ | 7/8 | 3 | 11 | 9 | 12 | 13 | 14 |
| $K$ | 8 | 3 | 11 | 9 | 12 | 13 | 14 |

From Table 8 we see that considered scientists are ranked with respect to the $B$-index ($B =: \sqrt{hg}$) very close as with respect to the $h$-index. The pairs of the rank of $h$-index and $B$-index of Moed, Van Raan, Martin, Narin and Garfield are (3, 4), (4, 3), (7, 8), (8, 9) and (9, 7), respectively.

Namely, nine scientists have the same rank with respect to these two indices. Martin and Narin have the same $h$-index, and they are ranked 7th - 8th place with respect to the $h$-index, and 8th and 9th place with respect to the $g$-index, respectively. Moreover, Garfield and Braun have the same $h$-index, and they are ranked 9th - 10th place with respect to the $h$-index, and 7th and 10th place with respect to the $g$-index, respectively. Moed and Van Raan are ranked 3th and 4th place with respect to the $h$-index, and 4th and 3th place with respect to the $g$-index, respectively.

**Remarks 3.10.** Table from Appendix A in Jin et al. (2007) gives the $h$-, $g$- and the $R$-indices of all 14 Price awardees, based on WoS, January 2006. Using this and the expression (21) of Corollary 2.15 given in the form

$$r = \left[\frac{2R^2}{h(h+1)}\right] - 1,$$

we obtain the following table.

**Table 9.** Price awardees - The $h$-, $g$-, $R$- and $H$-indices and the values $r$, $H/g$ and $H/R$

| Name | Leydesdorf | Glänzel | Moed | Van Raan | Rousseau | Schubert | Martin |
|---|---|---|---|---|---|---|---|
| $h$ | 13 | 18 | 18 | 19 | 13 | 18 | 16 |
| $g$ | 19 | 27 | 27 | 27 | 15 | 30 | 27 |
| $h/g$ | 0.684 | 0.667 | 0.667 | 0.704 | 0.867 | 0.600 | 0.593 |
| $R$ | 17.52 | 37.85 | 27.40 | 24.73 | 14.20 | 25.53 | 25.17 |
| $r$ | 2 | 2 | 3 | 2 | 1 | 2 | 3 |
| $H = h\sqrt{r}$ | 18.384 | 47.624 | 31.177 | 26.870 | 13 | 25.456 | 27.713 |
| $H/g$ | 0.968 | 1.764 | 1.155 | 0.995 | 0.867 | 0.849 | 1.026 |
| $H/R$ | 1.049 | 1.258 | 1.138 | 1.087 | 0.915 | 0.997 | 1.101 |

**Table 9** – Continued. Price awardees - The $h$-, $g$-, $R$- and $H$-indices and the values $r$, $H/g$ and $H/R$

| Name | Narin | Garfield | Braun | Small | Egghe | Ingwersen | White |
|---|---|---|---|---|---|---|---|
| $h$ | 27 | 27 | 25 | 18 | 13 | 13 | 12 |
| $g$ | 40 | 59 | 38 | 39 | 19 | 26 | 25 |
| $h/g$ | 0.675 | 0.458 | 0.658 | 0.462 | 0.684 | 0.500 | 0.480 |
| $R$ | 37.51 | 55.21 | 34.17 | 24.37 | 24.85 | 17.77 | 23.52 |
| $r$ | 2 | 7 | 2 | 2 | 5 | 2 | 6 |
| $H = h\sqrt{r}$ | 38.184 | 71.435 | 35.355 | 24.456 | 29.069 | 18.385 | 29.394 |
| $H/g$ | 0.955 | 1.211 | 0.930 | 0.653 | 1.530 | 0.707 | 1.176 |
| $H/R$ | 1.018 | 1.294 | 1.035 | 1.0035 | 1.170 | 1.035 | 1.250 |

From Table 9 we see that the values $H/g$ are less than 1 for eight Price awardees, while these values are close to 1 for five Price awardees (Leydesdorf, Van Raan, Martin, Narin and Braun, for which $|H/g-1| \leq 0.07$). Furthermore, the values $H/R$ are close to 1 for seven Price awardees (namely, we see $0 < H/g - 1 < 0.05$ for these seven Price awardees).

## 4. Conclusions

This paper presents new indices to characterize the scientific output of researchers. The $g_d$-indices $(d = 2,3,...,cit_1)$ were introduced as extensions of the Egghe's $g$-index whose definition allows defining the order $r = r(g,h)$ $(r = 1,2,...)$ of the $g$-index with respect to the Hirsch $h$-index. Further the $\overline{g}$-index is defined as $\overline{g} = g$ if $r = 1$, and $\overline{g}$ is equal to the average of the $g, g_3, ..., g_{r+1}$- indices if $r \geq 2$, and $H$-index as $H = h\sqrt{r}$. The definition of the order $r = r(g,h)$ suggests the fact that $H$-index and the $\overline{g}$-index significantly improve the $h$-index of scientists with highly cited publications which are not appropriately appreciated in the Hirsch $h$-index. This is confirmed by computational results concerning the values of these indices and related ratios for Price awardees. Notice that for all considered Price awardees it holds $h < \overline{g} < g < H < A$.

The $D$-, $B$-, $C$-, $K$-, and $F$-index are defined as the expressions involving the previous mentioned indices. We prove several inequalities involving these indices and earlier investigated the $h$-, $g$-, $A$- and $R$- and $hg$-indices. These inequalities and our computational results containing all these indices for 14 Price awardees, confirm very good approximations between some pairs of these indices. Finally, notice that certain indices defined in this paper eliminate some of the disadvantages of the $h$-index and $g$-index.

# References

Alonso, S., Cabrerizo, F.J, Herrera-Viedma, E. and Herrera, F. (2010). *hg*-index: A New Index to Characterize the Scientic Output of Researchers Based on the *h*- and g-indices, *Scientometrics*, *82*, 391–400.

Bihari, A., Tripathi, S. and Deepak, A. (2021). A review on *h*-index and its alternative indices, *Journal of Information Science*, 1–42, DOI: 10.1177/01655515211014478

Britoa, R. and Rodriguez Navarro, A. (2021). The inconsistency of *h*-index: A mathematical analysis, *Journal of Informetrics*, *15*(1), 101–106.

Egghe, L. (2006a). An improvement of the *h*-index: The *g*-index. *International Society for Scientometrics and Informetrics* (*ISSI newsletter*), *2*(1), 8–9.

Egghe, L. (2006b). How to improve the *h*-index: Letter. *The Scientist*, *20*(3).

Egghe, L. (2006c). Theory and practice of the *g*-index. *Scientometrics*, *69*(1), 131–152.

Egghe, L. (2010). The Hirsch index and related impact measures, *Information Science and Technology*, *44*(1), 65–114.

Egghe, L. and Rousseau, R. (2019). Solution by step functions of a minimum problem in $L^2[0,T]$, using generalized $h$ - and $g$ -indices, *Journal of Informetrics*, *13*(3), 785–792.

Glänzel, W. and Persson, O. (2005). *H*-index for Price medalist. ISSI Newsletter, *1*(4), 15–18

Hirsch, J. (2005). An index to quantify an individual's scientific research output. *Proceedings of the National Academy of Sciences of the United States of America – PNAS*, *102*, 16569–16572.

Jin, B.-H. (2007). The *AR*-index: Complementing the *h*-index, *ISSI Newsletter*, *3*(1), p. 6.

Jin, B.-H., Liang, L.-M., Rousseau, R. and Egghe, L. (2007). The *R*- and *AR*-indices: Complementing the *h*-index, *Chinese Science Bulletin*, *52*(6), 855–863.

Rousseau, R. (2006). New developments related to the Hirsch index, *Science Focus*, *1*(4), 23–25 (in Chinese). English version available at: E-LIS: code 6376, http://www.eprints.rclis.org/archive/00006376/.

Schreiber, M. (2010). Twenty Hirsch index variants and other indicators giving more or less preference to highly cited papers, *Annalen der Physik* (Berlin), *522*(8), 536-554.

van Eck, N. J. and Waltman, L. (2008). Generalizing the *h*- and *g*-indices. *Journal of Informetrics*, *2*(4), 263–271.

Waltman, L. (2016). A review of the literature on citation impact indicators. *Journal of Informetrics*, *10*(2), 365–391.

Woeginger J. G. (2009). Generalizations of Egghe's *g*-Index, *Journal of the American Society for Information Science and Technology*, *60*(6), 1267–1273.

# Appendix

This Appendix contains Price awardees data (see more in Glänzel and Persson, 2005) extracted from Scopus in Tables A1 and A2, their citation count per each paper and $h$-, $g$- and $g_d$-indices.

**Table A1.** Price awardees (see more in Glänzel and Persson, 2005) and their $h$-index and $g$-indices (bases on Scopus, 21$^{st}$ February 2023 and Jin et al. 2007)

| Scientist | No. of papers | No. of citations | No. of citations per paper | $h$-index | | $g$-index | |
|---|---|---|---|---|---|---|---|
| | | | | January 2006 WoS | 21 February 2023 Scopus | January 2006 WoS | 21 February 2023 Scopus |
| Eugene Garfield | 256 | 11515 | 45 | 27 | 37 | 59 | 106 |
| Francis Narin | 68 | 7209 | 106 | 27 | 38 | 40 | 68 |
| Tibor Braun | 260 | 5680 | 22 | 25 | 37 | 38 | 66 |
| Anthony Van Raan | 136 | 8308 | 61 | 19 | 48 | 27 | 89 |
| Wolfgang Glänzel | 293 | 11766 | 40 | 18 | 61 | 27 | 99 |
| Henk Moed | 147 | 7606 | 52 | 18 | 49 | 27 | 86 |
| Andras Schubert | 164 | 7587 | 46 | 18 | 41 | 30 | 85 |
| Henry Small | 62 | 7693 | 124 | 18 | 34 | 39 | 57 |
| Ben Martin | 96 | 7598 | 79 | 16 | 38 | 27 | 85 |
| Leo Egghe | 231 | 5640 | 24 | 13 | 30 | 19 | 69 |
| Peter Ingwersen | 100 | 3606 | 36 | 13 | 27 | 26 | 59 |
| Loet Leydesdorff | 432 | 25005 | 58 | 13 | 79 | 19 | 145 |
| Ronald Rousseau | 330 | 8053 | 24 | 13 | 43 | 15 | 79 |
| Howard White | 29 | 2399 | 83 | 12 | 19 | 25 | 28 |

Recall that in Table A2 the citations that belong to the bounds of the $h$-, $g$- and $g_d$-indices of 14 Price awardees are marked in bold. Also, recall that by Table 1, for Schubert $h = g_{r+1} = g_7 = 42$, for Narin $h = g_{r+1} = g_9 = 38$, for Braun $h = g_{r+1} = g_5 = 37$ and for Small $h = g_{r+1} = g_{12} = 34$ Notice also that by Table 1, for Narin $g = g_3 = 68$ (which is actually equal to the number of his publications), for Small $g = g_3 = g_4 = 57$ and for White $g = g_3 = g_4 = g_5 = 28$.

**Table A2.** Price awardees data and their $h$-, $g$- and $g_d$-indices (bases on Scopus, 21st February 2023)

| | Price awardees | | | | | | | | | | | | | |
|---|---|---|---|---|---|---|---|---|---|---|---|---|---|---|
| | Leyde-sdorff | Glänzel | Moed | Van Raan | Rousseau | Schubert | Martin | Narin | Garfield | Braun | Small | Egghe | Ingwe-Rsen | White |
| 1 | 3964 | 534 | 417 | 522 | 1050 | 1858 | 1828 | 803 | 1888 | 449 | 2846 | 1455 | 482 | 1029 |
| 2 | 647 | 449 | 417 | 455 | 484 | 449 | 728 | 780 | 1711 | 362 | 592 | 478 | 414 | 213 |
| 3 | 544 | 417 | 380 | 381 | 478 | 362 | 411 | 519 | 1635 | 255 | 540 | 242 | 348 | 178 |
| 4 | 479 | 364 | 350 | 350 | 265 | 273 | 391 | 481 | 597 | 158 | 410 | 206 | 232 | 148 |
| 5 | 424 | 273 | 303 | 303 | 242 | 255 | 371 | 443 | 589 | 151 | 242 | 153 | 166 | 124 |
| 6 | 422 | 255 | 290 | 290 | 175 | 211 | 362 | 354 | 430 | 147 | 197 | 124 | 156 | 123 |
| 7 | 410 | 249 | 288 | 284 | 153 | 159 | 282 | 339 | 264 | 138 | 193 | 110 | 155 | 104 |
| 8 | 401 | 220 | 248 | 283 | 126 | 158 | 244 | 306 | 255 | 135 | 190 | 107 | 133 | 72 |
| 9 | 376 | 211 | 197 | 276 | 118 | 155 | 223 | 283 | 252 | 127 | 186 | 88 | 113 | 54 |
| 10 | 352 | 201 | 183 | 206 | 111 | 151 | 196 | 226 | 238 | 120 | 157 | 81 | 70 | 50 |
| 11 | 352 | 188 | 182 | 172 | 110 | 147 | 179 | 193 | 238 | 119 | 154 | 71 | 57 | 35 |
| 12 | 301 | 163 | 174 | 166 | 97 | 133 | 131 | 183 | 178 | 96 | 151 | 70 | 56 | 32 |
| 13 | 291 | 159 | 166 | 157 | 96 | 127 | 128 | 159 | 155 | 77 | 144 | 70 | 49 | 29 |
| 14 | 266 | 155 | 152 | 152 | 88 | 120 | 122 | 151 | 154 | 75 | 129 | 68 | 46 | 28 |
| 15 | 252 | 147 | 146 | 150 | 87 | 119 | 114 | 138 | 153 | 73 | 113 | 65 | 45 | 25 |
| 16 | 242 | 145 | 122 | 146 | 75 | 105 | 101 | 121 | 151 | 69 | 112 | 50 | 42 | 24 |
| 17 | 241 | 141 | 118 | 142 | 74 | 98 | 77 | 114 | 139 | 67 | 111 | 48 | 42 | 23 |
| 18 | 232 | 135 | 118 | 136 | 72 | 92 | 73 | 94 | 132 | 61 | 102 | 47 | 41 | 21 |
| 19 | 225 | 133 | 117 | 130 | 70 | 89 | 72 | 87 | 114 | 59 | 98 | 46 | 39 | **20** |
| 20 | 221 | 127 | 103 | 111 | 67 | 75 | 69 | 81 | 113 | 57 | 90 | 45 | 39 | **19** |
| 21 | 184 | 126 | 95 | 110 | 66 | 73 | 68 | 76 | 111 | 55 | 81 | 42 | 38 | **17** |
| 22 | 179 | 126 | 92 | 103 | 64 | 73 | 66 | 73 | 91 | 55 | 79 | 41 | 35 | **15** |
| 23 | 171 | 121 | 88 | 98 | 63 | 72 | 65 | 73 | 88 | 52 | 71 | 40 | 35 | **5** |
| 24 | 166 | 111 | 80 | 94 | 61 | 71 | 64 | 68 | 83 | 52 | 57 | 40 | 33 | 5 |
| 25 | 165 | 109 | 76 | 92 | 58 | 70 | 63 | 67 | 74 | 51 | 51 | 39 | 29 | **2** |
| 26 | 157 | 108 | 75 | 86 | 56 | 69 | 58 | 61 | 73 | 47 | 49 | 35 | 29 | 2 |
| 27 | 154 | 96 | 75 | 82 | 56 | 69 | 57 | 61 | 69 | 46 | 47 | 34 | **28** | **1** |
| 28 | 147 | 94 | 73 | 81 | 55 | 64 | 55 | 55 | 65 | 45 | 44 | 34 | **27** | **1** |
| 29 | 147 | 92 | 71 | 81 | 51 | 63 | 50 | 54 | 64 | 45 | 43 | 34 | 27 | 0 |
| 30 | 147 | 89 | 70 | 77 | 49 | 62 | 48 | 50 | 63 | 45 | 41 | 32 | 26 | 0 |
| 31 | 141 | 89 | 69 | 76 | 48 | 62 | 48 | 43 | 63 | 43 | 40 | **30** | **25** | |
| 32 | 138 | 89 | 69 | 75 | 48 | 57 | 47 | 43 | 53 | 42 | 39 | 29 | 24 | |
| 33 | 135 | 85 | 63 | 73 | 47 | 53 | 46 | 43 | 51 | 40 | 38 | **29** | 24 | |
| 34 | 133 | 85 | 62 | 72 | 46 | 52 | 46 | 42 | 49 | 40 | 34 | 28 | **24** | |
| 35 | 133 | 84 | 61 | 72 | 45 | 47 | 45 | 41 | 45 | 39 | 21 | 27 | 23 | |
| 36 | 129 | 84 | 60 | 68 | 45 | 46 | 40 | 40 | 42 | 37 | **21** | **26** | 23 | |
| 37 | 129 | 83 | 60 | 65 | 45 | 46 | 40 | 40 | 39 | 37 | 20 | 25 | 23 | |
| 38 | 127 | 80 | 58 | 63 | 45 | 45 | 40 | 38 | **30** | 36 | **19** | 24 | 22 | |
| 39 | 127 | 80 | 57 | 62 | 44 | 45 | **32** | 37 | **28** | 34 | 18 | 24 | **22** | |
| 40 | 126 | 78 | 57 | 60 | 44 | 43 | 32 | 34 | 27 | 33 | **17** | 24 | 22 | |
| 41 | 125 | 76 | 56 | 59 | 43 | 41 | 31 | 32 | **27** | 33 | 17 | **24** | 18 | |
| 42 | 125 | 76 | 55 | 59 | 43 | **42** | 28 | **27** | 25 | **33** | 15 | 24 | 17 | |
| 43 | 125 | 75 | 53 | 59 | **43** | 40 | 24 | 24 | **23** | 32 | **14** | 23 | 16 | |
| 44 | 125 | 75 | 53 | 55 | 42 | 37 | 23 | **21** | 22 | 31 | 10 | 22 | 16 | |
| 45 | 124 | 75 | 52 | 52 | **42** | 36 | **23** | 21 | **22** | 31 | 9 | 22 | 15 | |
| 46 | 123 | 74 | 52 | 51 | 42 | **36** | 23 | 20 | 21 | 30 | **9** | 22 | **15** | |
| 47 | 122 | 73 | 50 | 48 | 41 | 36 | 22 | 18 | 21 | 29 | 6 | 22 | 14 | |
| 48 | 122 | 73 | 49 | **48** | 40 | 33 | **21** | **18** | **21** | 28 | 4 | 22 | 14 | |
| 49 | 118 | 73 | **49** | 47 | 39 | 32 | 21 | 17 | 21 | 28 | 4 | 20 | 13 | |
| 50 | 108 | 72 | **47** | 45 | 39 | 31 | 21 | 14 | 21 | 27 | **4** | 20 | 13 | |
| 51 | 106 | 72 | 45 | 41 | 39 | **31** | 20 | 12 | **20** | **27** | 4 | 20 | 13 | |
| 52 | 102 | 71 | 44 | **41** | **37** | 27 | 20 | 12 | 20 | 27 | 2 | 19 | 12 | |
| 53 | 100 | 71 | 43 | 39 | 37 | 27 | 20 | **11** | 20 | 27 | 1 | 19 | 12 | |
| 54 | 98 | 70 | 43 | 39 | 33 | 27 | **17** | 11 | 19 | 26 | **1** | **18** | 11 | |
| 55 | 97 | 69 | 41 | 39 | 32 | 27 | 16 | 11 | **18** | 26 | 1 | 18 | 10 | |
| 56 | 94 | 66 | 41 | 38 | 32 | 26 | 14 | 9 | 18 | 26 | 1 | 18 | 10 | |
| 57 | 94 | 64 | **39** | 38 | 30 | 26 | 11 | 7 | 17 | 25 | **1** | 18 | 9 | |
| 58 | 93 | 63 | 39 | 37 | 30 | **25** | 9 | 6 | 17 | 25 | | 18 | 8 | |
| 59 | 93 | 62 | 36 | **37** | 29 | 23 | 9 | **6** | 17 | 25 | | **17** | 8 | |
| 60 | 92 | 62 | 35 | 36 | 29 | 23 | **9** | 6 | **16** | 25 | | 17 | 8 | |
| 61 | 91 | **62** | 34 | 36 | 28 | 23 | 8 | 5 | 16 | 25 | | 17 | 8 | |

**Table A2** – Continued. Price awardees data and their $h$-, $g$- and $g_d$-indices (bases on Scopus, 21st February 2023)

| | Price awardees | | | | | | | | | | | | | |
|---|---|---|---|---|---|---|---|---|---|---|---|---|---|---|
| | Leyde-sdorff | Glänzel | Moed | Van Raan | Rousseau | Schubert | Martin | Narin | Garfield | Braun | Small | Egghe | Ingwe-Rsen | White |
| 62 | 90 | 59 | 33 | 35 | **28** | 23 | 7 | 4 | 15 | 24 | | 17 | 7 | |
| 63 | 89 | **59** | 32 | 34 | 28 | 22 | 7 | 2 | 15 | 24 | | 16 | 7 | |
| 64 | 87 | 57 | 30 | 32 | 27 | 22 | 7 | 1 | 14 | 24 | | 16 | 6 | |
| 65 | 87 | 57 | 29 | 31 | 27 | 21 | 7 | 0 | **14** | 23 | | 16 | 5 | |
| 66 | 87 | 56 | 29 | 31 | 27 | 20 | 6 | 0 | 14 | **23** | | 16 | 5 | |
| 67 | 87 | 55 | 28 | 30 | 27 | 19 | 6 | 0 | 14 | 23 | | 16 | 4 | |
| 68 | 86 | 54 | **27** | 29 | 27 | **17** | 5 | **0** | 14 | 22 | | 15 | 4 | |
| 69 | 83 | 54 | 27 | 29 | 26 | 17 | 5 | | 13 | 22 | | **14** | 4 | |
| 70 | 83 | 53 | 27 | **28** | 26 | 16 | 5 | | 13 | 21 | | 14 | 4 | |
| 71 | 83 | 52 | 25 | 28 | 25 | 15 | 5 | | 12 | 21 | | 14 | 4 | |
| 72 | 82 | 52 | 25 | 28 | 24 | 15 | 4 | | 12 | 21 | | 14 | 3 | |
| 73 | 82 | 52 | 23 | 27 | 24 | 15 | 4 | | 12 | 21 | | 14 | 3 | |
| 74 | 82 | 52 | 23 | 25 | 24 | 15 | 4 | | **12** | 20 | | 14 | 3 | |
| 75 | 80 | 51 | 22 | 24 | 24 | 14 | 4 | | 12 | 19 | | 14 | 3 | |
| 76 | 80 | **49** | 22 | 24 | 23 | 14 | 4 | | 12 | 19 | | 13 | 3 | |
| 77 | 80 | 48 | 22 | 24 | 23 | 14 | 3 | | 10 | 19 | | 13 | 2 | |
| 78 | 79 | 48 | 19 | 23 | 23 | 14 | 3 | | 10 | 18 | | 13 | 2 | |
| 79 | 79 | 48 | 18 | 22 | **23** | 13 | **2** | | 10 | 18 | | 13 | 2 | |
| 80 | 78 | 47 | 16 | 21 | 23 | 13 | 2 | | 9 | 17 | | 12 | 2 | |
| 81 | 78 | 46 | 16 | 21 | 23 | 13 | 1 | | 9 | 17 | | 12 | 1 | |
| 82 | 77 | 45 | 15 | 20 | 23 | 13 | 1 | | 9 | 16 | | 12 | 1 | |
| 83 | **77** | 45 | 14 | 20 | 22 | 13 | 1 | | 9 | 16 | | 11 | 1 | |
| 84 | 77 | 44 | 13 | 20 | 22 | 12 | 1 | | 8 | 16 | | 11 | 1 | |
| 85 | 76 | 44 | 13 | 20 | 21 | **12** | **1** | | 7 | 16 | | 11 | 1 | |
| 86 | 76 | 43 | **12** | 18 | 21 | 12 | | | **7** | 16 | | 11 | 1 | |
| 87 | 76 | 43 | 11 | 18 | 21 | 12 | | | 7 | 16 | | 11 | 1 | |
| 88 | 75 | 42 | 11 | 17 | 21 | 10 | | | 6 | 16 | | 11 | 1 | |
| 89 | 75 | 41 | 10 | **17** | 20 | 10 | | | 6 | 15 | | 11 | | |
| 90 | 74 | 40 | 10 | 17 | 20 | 9 | | | 6 | 15 | | 10 | | |
| 91 | 74 | 38 | 9 | 17 | 20 | 9 | | | 6 | 15 | | 10 | | |
| 92 | 73 | 38 | 9 | 13 | 20 | 9 | | | 6 | 15 | | 10 | | |
| 93 | 71 | 38 | 9 | 13 | 20 | 9 | | | 5 | 15 | | 10 | | |
| 94 | 70 | 36 | 8 | 13 | 19 | 8 | | | 5 | 15 | | 10 | | |
| 95 | 69 | 36 | 8 | 13 | 19 | 8 | | | 5 | 15 | | 10 | | |
| 96 | **68** | 36 | 8 | 12 | 18 | 8 | | | 5 | 14 | | 10 | | |
| 97 | 68 | 35 | 7 | 11 | 18 | 7 | | | 5 | 14 | | 10 | | |
| 98 | 68 | 34 | 7 | 10 | 18 | 7 | | | 5 | 14 | | 10 | | |
| 99 | 67 | **34** | 7 | 10 | 18 | 6 | | | 5 | 14 | | 10 | | |
| 100 | 66 | 34 | 7 | 10 | 17 | 6 | | | 5 | 14 | | 10 | | |
| 101 | 65 | 33 | 7 | 9 | 17 | 6 | | | 5 | 14 | | 9 | | |
| 102 | 64 | 33 | 6 | 9 | 17 | 6 | | | 5 | 14 | | 9 | | |
| 103 | 64 | 33 | 6 | 9 | 17 | 6 | | | 5 | 14 | | 9 | | |
| 104 | 63 | 32 | 6 | 9 | 17 | 6 | | | 4 | 14 | | 9 | | |
| 105 | 63 | 32 | 5 | 8 | 17 | 5 | | | 4 | 13 | | 9 | | |
| 106 | 62 | 32 | 5 | 8 | 16 | 5 | | | **4** | 13 | | 9 | | |
| 107 | 61 | 31 | 5 | 8 | 16 | 5 | | | | 13 | | 9 | | |
| 108 | 61 | 31 | 4 | 8 | 16 | 4 | | | | 13 | | 9 | | |
| 109 | 60 | 31 | 4 | 7 | 16 | 4 | | | | 13 | | 9 | | |
| 110 | 60 | 30 | 4 | 7 | 16 | 4 | | | | 13 | | 9 | | |
| 111 | 59 | 30 | 4 | 6 | 16 | 4 | | | | 13 | | 9 | | |
| 112 | 59 | 30 | 4 | 5 | 15 | 4 | | | | 13 | | 9 | | |
| 113 | 58 | 29 | 3 | 5 | 15 | 3 | | | | 13 | | 8 | | |
| 114 | **58** | 28 | 3 | 4 | 15 | 3 | | | | 12 | | 8 | | |
| 115 | 56 | 27 | 3 | 4 | 15 | 3 | | | | 12 | | 8 | | |
| 116 | 56 | 27 | 2 | 4 | 14 | 3 | | | | 12 | | 8 | | |
| 117 | 56 | 27 | 2 | 3 | 14 | 2 | | | | 12 | | 8 | | |
| 118 | 55 | 27 | 2 | 3 | 14 | 2 | | | | 12 | | 8 | | |
| 119 | 55 | 27 | 1 | 3 | 14 | 2 | | | | 12 | | 7 | | |
| 120 | 55 | 27 | 1 | 2 | 14 | 2 | | | | 11 | | 7 | | |
| 121 | 55 | 27 | 1 | 2 | 13 | 2 | | | | 11 | | 7 | | |
| 122 | 53 | 26 | 1 | 1 | 13 | 2 | | | | 10 | | 7 | | |

**Table A2** – Continued. Price awardees data and their $h$-, $g$- and $g_d$-indices (bases on Scopus, 21st February 2023)

| | Price awardees | | | | | | | | | | | | | |
|---|---|---|---|---|---|---|---|---|---|---|---|---|---|---|
| | Leyde-sdorff | Glänzel | Moed | Van Raan | Rousseau | Schubert | Martin | Narin | Garfield | Braun | Small | Egghe | Ingwe-rsen | White |
| 123 | 52 | 26 | 1 | 1 | 13 | 2 | | | | 10 | | 7 | | |
| 124 | 51 | 26 | 1 | 1 | 13 | 2 | | | | 10 | | 7 | | |
| 125 | 51 | 26 | 1 | | 13 | 2 | | | | 10 | | 6 | | |
| 126 | 49 | 26 | 1 | | 13 | 2 | | | | 10 | | 6 | | |
| 127 | 49 | 25 | 1 | | 13 | 2 | | | | 9 | | 6 | | |
| 128 | 48 | 25 | | | 13 | 2 | | | | 9 | | 6 | | |
| 129 | 47 | 24 | | | 13 | 1 | | | | 9 | | 6 | | |
| 130 | 46 | 24 | | | 13 | 1 | | | | 9 | | 6 | | |
| 131 | 44 | 24 | | | 13 | 1 | | | | 9 | | 5 | | |
| 132 | 44 | 23 | | | 13 | 1 | | | | 9 | | 5 | | |
| 133 | 44 | 23 | | | 12 | 1 | | | | 9 | | 5 | | |
| 134 | 44 | 23 | | | 12 | 1 | | | | 9 | | 5 | | |
| 135 | 43 | 23 | | | 12 | 1 | | | | 9 | | 5 | | |
| 136 | 43 | 23 | | | 12 | 1 | | | | 8 | | 5 | | |
| 137 | 43 | 22 | | | 11 | 1 | | | | 8 | | 5 | | |
| 138 | 43 | 21 | | | 11 | 1 | | | | 8 | | 5 | | |
| 139 | 43 | 21 | | | 11 | 1 | | | | 8 | | 5 | | |
| 140 | 43 | 21 | | | 11 | 1 | | | | 8 | | 5 | | |
| 141 | 42 | 20 | | | 11 | 1 | | | | 8 | | 5 | | |
| 142 | 42 | 19 | | | 11 | 1 | | | | 8 | | 5 | | |
| 143 | 41 | 19 | | | 11 | | | | | 8 | | 5 | | |
| 144 | 41 | 18 | | | 11 | | | | | 8 | | 4 | | |
| 145 | 41 | 18 | | | 10 | | | | | 7 | | 4 | | |

Note:

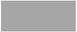 $h$-index      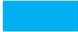 $g$-index      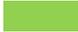 $g_3, g_4, g_5,\ldots, g_{14}$-indices